\documentclass[useAMS,usenatbib,usegraphicx]{mn2e}

\newcommand{\etal}{{et al.\ }}

\title[Narrow-Band Transit Photometry with GTC/OSIRIS]{Characterizing Transiting Extrasolar Planets with Narrow-Band Photometry and GTC/OSIRIS}

\author[K. D. Col\'on \etal]{Knicole D. Col\'on$^{1}$\thanks{E-mail: knicole@astro.ufl.edu}\thanks{NSF Graduate Research Fellow.}, Eric B.\ Ford$^{1}$, Brian Lee$^{1}$, Suvrath Mahadevan$^{2,3}$, 
\newauthor
Cullen H. Blake$^{4}$\thanks{NSF Astronomy \& Astrophysics Postdoctoral Fellow.}\\
$^{1}$Department of Astronomy, University of Florida, Gainesville, FL 32611, USA\\
$^{2}$Department of Astronomy \& Astrophysics, Pennsylvania State University, University Park, PA 16802, USA\\
$^{3}$Center for Exoplanets and Habitable Worlds, Pennsylvania State University, University Park, PA 16802, USA\\
$^{4}$Department of Astrophysical Sciences, Princeton University, Princeton, NJ 08544, USA}

\begin{document}

\date{Accepted 2010 June 16. Received 2010 June 14; in original form 2010 February 10}

\pagerange{\pageref{firstpage}--\pageref{lastpage}} \pubyear{2010}

\maketitle

\label{firstpage}

\begin{abstract} 
We report the first extrasolar planet observations from the 10.4-m Gran Telescopio Canarias (GTC), currently the world's largest, fully steerable, single-aperture optical telescope.  We used the OSIRIS tunable filter imager on the GTC to acquire high-precision, narrow-band photometry of the transits of the giant exoplanets, TrES-2b and TrES-3b.  We obtained near-simultaneous observations in two near-infrared (NIR) wavebands (790.2 and 794.4$\pm$2.0 nm) specifically chosen to avoid water vapor absorption and skyglow so as to minimize the atmospheric effects that often limit the precision of ground-based photometry.  Our results demonstrate a very-high photometric precision with minimal atmospheric contamination despite relatively poor atmospheric conditions and some technical problems with the telescope.  We find the photometric precision for the TrES-2 observations to be 0.343 and 0.412 mmag for the 790.2 and 794.4 nm light curves, and the precision of the TrES-3 observations was found to be 0.470 and 0.424 mmag for the 790.2 and 794.4 nm light curves.  We also discuss how future follow-up observations of transiting planets with this novel technique can contribute to the characterization of Neptune- and super-Earth-size planets to be discovered by space-based missions like {\em CoRoT} and {\em Kepler}, as well as measure atmospheric properties of giant planets, such as the strength of atmospheric absorption features.
\end{abstract}

\section{Introduction} 
\label{SecIntro}

Space-based missions like {\em CoRoT} and {\em Kepler} have the photometric capability to detect transiting planets with radii not much larger than the Earth's, commonly referred to as ``super-Earths'' \citep{borucki2009,leger2009}. While space-based observatories have provided the highest photometric precisions, ground-based follow-up observations play an essential role in confirming detections of transiting planets and characterizing the planets' orbits, interiors, compositions and atmospheres \citep{torres2008,johnson2009}.  In particular, because {\em CoRoT} and {\em Kepler}'s white light observations are affected by stellar limb darkening (LD), there can be strong correlations between the stellar LD parameters, the transit impact parameter, and the transit duration (which depends on the eccentricity and pericentre direction).  High-precision photometry conducted at far-red, near-infrared (NIR) or mid-infrared wavelengths can break this degeneracy, providing more precise measurements of a planet's orbital and physical parameters \citep{colon2009}.  Observing at these wavelengths also allows for reduced differential atmospheric effects, which can further improve the quality of the transit light curve.

The detection of transiting Earth-size planets around solar-like stars requires such high photometric precision that astronomers had long assumed that characterizing such planets could only be done from space to avoid the deleterious effects of Earth's atmosphere \citep{borucki85}.  Narrow-band photometry with large telescopes provides an alternative path towards high photometric precision.  While inefficient for planet searches, near-simultaneous narrow-band observations provide additional opportunities for the characterization of known planets (or previously identified planet candidates), such as the measurement of atmospheric absorption.

In this paper we describe a novel observational technique for high-precision transit photometry, using near simultaneous observations in multiple narrow bandpasses (\S \ref{SecObs}).  We describe our data reduction in \S \ref{SecReduction} and light curve analysis in \S \ref{SecAnalysis}.  In \S \ref{SecResults} we present results for TrES-2 and TrES-3 and demonstrate that this technique can achieve a very high photometric precision.  Finally, in \S \ref{SecDiscuss}, we compare our results to those found in the literature, and we discuss the implications for studying the atmospheric composition of giant planets and for characterizing the bulk properties of super-Earth-size planets, including those in the habitable zone of main-sequence stars.

\section{Observations}
\label{SecObs}

The 10.4-m Gran Telescopio Canarias (GTC)\footnote[1]{http://www.gtc.iac.es/en/} is located at the Observatorio del Roque de los Muchachos, on the island of La Palma.  While the GTC began scientific operations in March of 2009, commissioning of the telescope and first light instruments is ongoing.  We describe some of the first scientific observations with the GTC using the Optical System for Imaging and low Resolution Integrated Spectroscopy (OSIRIS) in the tunable filter (TF) imaging mode \citep{cepa2000,cepa2003}.  OSIRIS includes two 2048 $\times$ 4096 pixel E2V 44-82 CCDs which provide a maximum unvignetted field of view of 7.8 $\times$ 7.8 arcmin.  OSIRIS offers a suite of filters, including a TF which allows the user to specify custom bandpasses with a central wavelength of 651-934.5 nm and a FWHM of 1.2-2.0 nm.  Note that the TF bandpass is not uniform across the field of view, with the effective wavelength decreasing radially outward from the centre.  Therefore, for the wavelengths used in these observations, a difference of $\sim$10 nm exists between the tuned wavelength at the optical centre and the wavelength observed at 4 arcmin from the optical centre (i.e. near one edge of the CCD).

We observed one transit each of two giant extrasolar planets, TrES-2b and TrES-3b, using GTC/OSIRIS.  The field of view was chosen so that: 1) the target and a ``primary'' reference star were observed at the same wavelength using the same CCD, and 2) several additional ``secondary'' reference stars were observed with the same CCD but at different distances from the optical centre (and thus at different wavelengths).  Due to the restrictions in CCD orientation, different chips in the OSIRIS CCD mosaic were used for the TrES-2 and TrES-3 observations.

During each transit, observations alternated between two bandpasses centred on 790.2 and 794.4 nm at the location of the target.  These wavebands were chosen to minimize atmospheric effects by minimizing water vapor absorption and skyglow.  Each observation was followed by 33.4-s of dead time for readout.  We use 1$\times$1 binning but utilize a fast readout mode (500 kHz) in order to decrease the dead time between exposures.  Recent and future CCD controller upgrades will reduce the dead time between exposures.  The telescope was also defocused to increase efficiency and to reduce the impact of pixel-to-pixel sensitivity variations.  Even with a slight defocus, the resulting point-spread functions of the stars were fairly well defined (i.e., not doughnut-shaped).

The TrES-2 ($V \sim$11.4) observations took place under photometric conditions on 2009 June 25 (dark time) from 1:58 to 5:43 UT, during which the airmass ranged from $\sim$1.07 to 1.44.  The defocussed FWHM of the target varied from $\sim$1.6-2.8 arcsec ($\sim$12.5-22 pixels) during the observations, while the actual seeing was $\sim$1.2 arcsec.  The autoguider system kept the images aligned within a few pixels, with the target's centroid coordinates shifting by less than 4 pixels in the x-direction and less than 6 pixels in the y-direction.  We used 80-s exposures, resulting in an overall cadence of 3.78 min for each bandpass.  Four of the images were excluded from our analysis due to tracking problems.  The last $\sim$16 min of out-of-transit (OOT) data were affected by twilight and are not included in the analysis.

The TrES-3 ($V \sim$12.4) observations took place under photometric conditions during grey time, starting at 23:53 UT on 2009 August 10 and continuing until 2:54 UT the next morning.  During the observations, the airmass ranged from $\sim$1.13 to 2.22, the defocussed FWHM of the target varied from $\sim$1.4-2.1 arcsec ($\sim$11-16.5 pixels), and the target's centroid coordinates shifted by less than 3 pixels in the x-direction and less than 5 pixels in the y-direction.  The exposure time was 120-s, yielding a 5.11 min observing cadence for each bandpass.  Problems with the telescope caused our observations to begin shortly after ingress had already begun.

\section{Data Reduction}
\label{SecReduction}

Standard IRAF procedures for bias subtraction and flat field correction were used for the TrES-2 observations.  However, during preliminary analysis, we found that the total number of counts in the dome flat fields (with a given exposure time) were decreasing with time.  Thus, for flat fielding, we include only those dome flats taken after the first 30 min of the lamp being turned on, by which time the lamp intensity had stabilized.  For TrES-2, we combine 40 (out of 60) dome flats for each observed wavelength (790.2 and 794.4 nm).  After summing over either set of flats, the average number of counts per pixel was 1.71$\times$10$^6$.  For the TrES-3 observations, we perform similar reductions as for TrES-2, but we only use 45 of the 95 total dome flats taken for each wavelength.  After summing over multiple flats, the average number of counts per pixel was 5.28$\times$10$^5$ (1.04$\times$10$^6$) for 790.2 (794.4) nm.  For the TrES-3 data reduction, we replace the median bias frame with the median of a series of dark frames taken with the same exposure time as the observations (albeit taken several months after the actual observations took place), since early observations revealed a higher than expected dark current.  We note that subtracting darks was beneficial only for the TrES-3 data, based on the RMS scatter of the out-of-transit light curve for TrES-3.  The opposite is true for TrES-2; i.e., the RMS scatter of the out-of-transit light curve was smaller when bias frames were subtracted (instead of dark frames).  This difference is likely due to the fact that the dark frames for both targets were taken in October 2009, rather than the exact same conditions as the transit observations (i.e., several months after each transit was observed).  Therefore, each data set was affected differently by its respective set of dark frames, resulting in the darks proving to be useful only for the TrES-3 observations. Recent hardware upgrades have reduced the dark current.

Due to the TF's small bandpass and position-dependent wavelength, all observations contain sky (OH) emission rings.  Therefore, we performed sky subtraction on all images based on the IRAF package TFred\footnote[2]{Written by D. H. Jones for the Taurus Tunable Filter, previously installed on the Anglo-Australian Telescope; http://www.aao.gov.au/local/www/jbh/ttf/adv\_reduc.html}, which estimates the sky background, including rings due to sky emission.  We performed aperture photometry on each target and reference star using the standard IDL routine APER\footnote[3]{http://idlastro.gsfc.nasa.gov/}.  We repeated this analysis using a range of aperture radii, and measured the root mean square (RMS) scatter of the flux ratio (target over sum of references) outside of transit in each colour.  After considering the results in both bandpasses, we adopted an aperture radius of 44 pixels (5.6 arcsec) for stars in the field of TrES-2 and 38 pixels (4.8 arcsec) for stars in the field of TrES-3.   Note that the use of TFred removes the need for a sky annulus, so we did not input one in the APER routine. 

To account for any atmospheric extinction, we fit linear airmass trends to each reference star's light curve (LC; computed by dividing the flux for a given star by the sum of the flux from all the other reference stars).  For each reference star, we discarded points that resulted in a flux ratio greater than 3$\sigma$ from the mean (typically one and at most two points per reference star).  We computed the reference flux at each time as the weighted sum of the flux of the remaining reference stars.  The flux in the target aperture was divided by the reference to compute the final LCs.  We found that using an ensemble of six to eight reference stars resulted in a smaller RMS OOT scatter than using just the primary reference, despite the fact that the secondary references were observed at different wavelengths and that the reference stars in the field of TrES-2 (TrES-3) ranged over $\sim$2 (3) magnitudes in brightness.  Since most of the reference stars were fainter than the targets, we conclude that adding additional faint references helps to reduce the photon noise for a larger ensemble and balance stellar variability within different reference stars.  
  
The uncertainties in the flux ratio were estimated by computing the quadrature sum of the photon noise for the target and the reference ensemble [median values are 0.331 (0.339) and 0.244 (0.249) mmag for TrES-2 and 0.426 (0.433) and 0.210 (0.214) mmag for TrES-3 for 790.2 (794.4) nm], the uncertainty in the sum of sky background and dark current for the target and the reference ensemble [median values are 0.0751 (0.0737) and 0.103 (0.105) mmag for TrES-2 and 0.0998 (0.0840) and 0.0688 (0.0667) mmag for TrES-3 for 790.2 (794.4) nm], and the scintillation noise for the target and primary reference [median values are 0.0627 (0.0630) mmag for TrES-2 and 0.0778 (0.0766) mmag for TrES-3 for 790.2 (794.4) nm].  The uncertainty in the sum of sky background and dark current was estimated by performing aperture photometry on the sky frames produced by TFred at the specific location of each target and reference star.  The scintillation noise was estimated using the relation given by \citet{dravins1998}, based on \citet{young67}.  We caution that this is an empirical relation and has not been tested for large telescopes at excellent sites like La Palma.  Thus, we consider the expression for scintillation noise to be only a rough estimate.  Nevertheless, it demonstrates that scintillation is expected to be only a very small contribution to the total error budget, thanks to the narrow filter bandpass.  Readout and digitization noise as well as flat field noise are negligible compared to photon noise and are not included in the calculation of the uncertainties in the flux ratio, so we include estimates for these noise sources \citep[as based on relations given by][]{southworth2009} only for reference.  The median readout noise is 7.44 electrons per pixel or 0.0116 (0.0123) mmag for TrES-2 and 7.35 electrons per pixel or 0.0150 (0.0158) mmag for TrES-3 for 790.2 (794.4) nm.  The median flat field noise is 6.07e-4 (6.07e-4) electrons per pixel or 0.00274 (0.00269) mmag for TrES-2 and 1.07e-3 (7.63e-4) electrons per pixel or 0.00778 (0.00535) mmag for TrES-3 for 790.2 (794.4) nm.  Based on the relation given by \citet{howell2006}, we find the median total uncertainties in the flux ratio for each exposure to be 0.434 (0.443) mmag for TrES-2 and 0.495 (0.499) mmag for TrES-3 for 790.2 (794.4) nm.  We note that our estimated uncertainties are somewhat larger than the measured residual RMS given in \S\ref{SecBins}. While the estimated and measured precision are marginally consistent, it is possible that we overestimated the uncertainty in individual measurements.

\begin{figure}
\includegraphics[width=84mm]{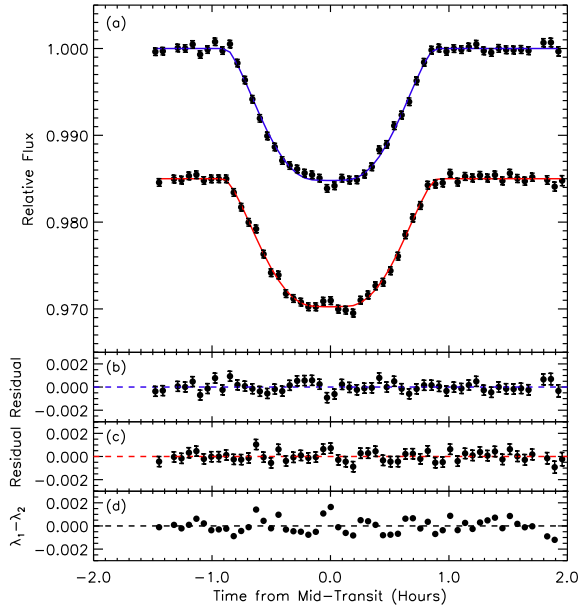}
\caption{Normalized light curves (a), residuals (b,c), and colour (d) for nearly simultaneous observations at 790.2 and 794.4$\pm$2.0 nm of TrES-2 as observed on UT 2009 June 25.  The filled circles are observations and the lines in panel (a) show the best-fitting models.  In panel (a), the 794.4-nm light curve has been arbitrarily offset by 0.015.  Panels (b) and (c) show residuals from the fits for the 790.2 and 794.4 nm light curve.   The colour of the residuals is shown in panel (d).}
\label{FigTres2}
\end{figure}

\begin{figure}
\includegraphics[width=84mm]{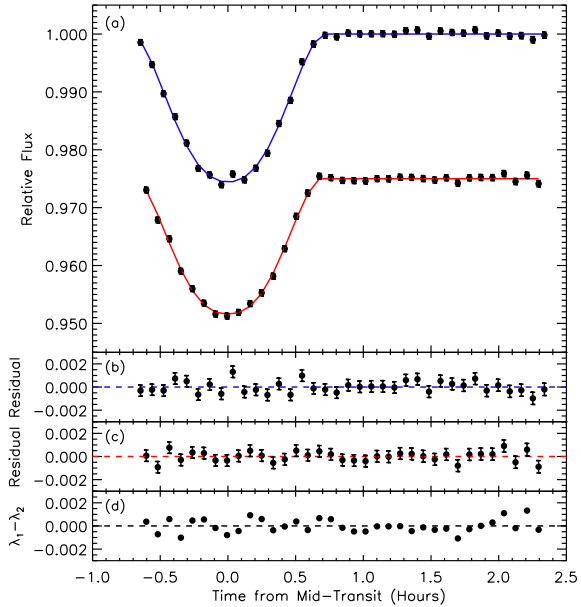} 
\caption{Same as in Figure \ref{FigTres2}, but for TrES-3 as observed on UT 2009 August 10.  In panel (a), the 794.4-nm light curve has been arbitrarily offset by 0.025.}
\label{FigTres3}
\end{figure}

We perform the above analysis for both bandpasses and each target.  The resulting photometric time series is reported and shown (with some corrections; see \S \ref{SecResults}) in Table \ref{tab_one} (full version available online) and Figures \ref{FigTres2} and \ref{FigTres3}.

\begin{table}
  \caption{Relative Photometry of TrES-2 and TrES-3 \label{tab_one} }
  \begin{tabular}{@{}cccc@{}}
  \hline
Observed $\lambda$ & HJD & Relative Flux & Uncertainty \\
\hline
\multicolumn{4}{c}{TrES-2} \\
790.2nm & 2455007.5845 & 0.99963 & 0.00043 \\
790.2nm & 2455007.5871 & 0.99969 & 0.00043 \\
790.2nm & 2455007.5923 & 1.00006 & 0.00043 \\
\multicolumn{4}{c}{\dots} \\
794.4nm & 2455007.5858 & 0.99956 & 0.00044 \\
794.4nm & 2455007.5910 & 0.99996 & 0.00044 \\
794.4nm & 2455007.5936 & 0.99984 & 0.00044 \\
\multicolumn{4}{c}{\dots} \\
\multicolumn{4}{c}{} \\
\multicolumn{4}{c}{TrES-3} \\
790.2nm & 2455054.4976 & 0.99855 & 0.00048 \\
790.2nm & 2455054.5011 & 0.99472 & 0.00048 \\
790.2nm & 2455054.5047 & 0.98969 & 0.00049 \\
\multicolumn{4}{c}{\dots} \\
794.4nm & 2455054.4993 & 0.99800 & 0.00049 \\
794.4nm & 2455054.5029 & 0.99266 & 0.00049 \\
794.4nm & 2455054.5064 & 0.98936 & 0.00049 \\
\multicolumn{4}{c}{\dots} \\
\hline
\end{tabular}

\medskip
The time stamps included here are for the times at mid-exposure, and the relative flux has been corrected for slopes/trends in the light curves (see \S\ref{SecResults} for more details).

\end{table}

\section{Light Curve Analysis}
\label{SecAnalysis}

Before fitting models to the light curves, we first applied the external parameter decorrelation (EPD) technique \citep[see, e.g.,][]{bakos2007,bakos2010} to each light curve in order to remove any systematic trends that are correlated with the following parameters: the X and Y centroid coordinates of the target on the image frames, the sharpness of the target's profile [approximately equal to (2.35/FWHM)$^2$] and the airmass.  We then performed the following analysis on the decorrelated light curves.

We fit the flux ratio with a standard planet transit model that includes a quadratic LD law \citep{mandel2002}.  We parametrize the LC model using: the time of mid-transit ($t_0$), impact parameter ($b\equiv a \cos i/R_{\star}$), transit duration (from first to fourth contact; $D$), planet-star radius ratio ($p \equiv R_p/R_{\star}$), average OOT flux ratio, a linear slope for the OOT flux ratio ($\alpha$), and two LD coefficients ($c_1 \equiv u_1 + u_2$ and $c_2 \equiv u_1 - u_2$), where $u_1$ and $u_2$ are the linear and quadratic LD coefficients of \citet{mandel2002}.

We used the publicly available code $mpfitfun$\footnote[4]{http://www.physics.wisc.edu/$\sim$craigm/idl/idl.html} to perform Levenberg-Marquardt minimization of $\chi^2$ to identify a best-fitting model for the transit photometry.  The initial guesses for $b$, $D$ and $p$ are based on estimates (and their uncertainties) from \citet{holman2007} and \citet{sozzetti2007} for TrES-2b and \citet{gibson2009} and \citet{sozzetti2009} for TrES-3b.  The initial guesses for $t_0$ were based on the ephemeris of \citet{rabus2009} for TrES-2 and \citet{sozzetti2009} for TrES-3.  In each case described below, we repeat the local minimization using an array of initial guesses for the modeled parameters based on the published uncertainties and concluded the results of the non-linear model fitting were not sensitive to our initial guesses.

Similarly, we tested several sets of values for $c_1$ and $c_2$ based on the theoretical LD coefficients estimated for each star.  Specifically, both LD coefficients were computed for our specific bandpasses for a grid of stellar models using PHOEBE \citep{prsa2005}.  We interpolated in ($T_{eff}$, $\log g$, [Fe/H]) to estimate $c_1$ and $c_2$ for each of several sets of stellar parameters.  We adopted stellar parameters from \citet{sozzetti2007} for TrES-2 and from \citet{sozzetti2009} for TrES-3, but we note that these abundance measurements are systematically lower than those determined by \citet{ammleretal2009}.  \citet{fischer2005} emphasize that different studies can yield different results due to, e.g., the analysis techniques and/or stellar models used, but both \citet{sozzetti2007,sozzetti2009} and \citet{ammleretal2009} use very similar techniques to estimate stellar parameters.  Therefore, we chose to interpolate LD coefficients over the range of the stellar parameters plus their measurement uncertainties as given by \citet{sozzetti2007,sozzetti2009}, since their analysis used slightly higher resolution spectra than used by \citet{ammleretal2009}.  As an extra precaution, we confirmed that their measurement uncertainties were realistic by comparing them to the adopted uncertainties given by \citet{fischer2005}, which were based on the analysis of observations of the solar spectrum as reflected by the asteroid Vesta.

Most previous studies use two parameter limb darkening laws and fit for one or both limb darkening parameters.  \citet{Southworth2008} investigated the effects of various limb darkening models and assumptions.  He concluded that a linear limb darkening law was often inadequate, so a non-linear law should be used, but that two parameter fits were typically highly degenerate.  In the cases of TrES-2 and TrES-3, two-parameter limb darkening models are particularly degenerate, since both have a large impact parameter and the planet only probes the stellar surface brightness near the stellar limb.  \citet{Southworth2008} also found that including uncertainty in limb darkening parameters was important to obtain realistic uncertainties for transit model parameters.  Therefore, \citet{Southworth2008} recommended holding $c_2$ fixed and fitting for $c_1$.  Initially, we tried this approach using separate limb darkening parameters for each star and each bandpass and fit for $c_1$.  We found that the best-fitting limb darkening coefficients for one star and our two bandpasses could differ significantly, even though stellar atmosphere models predict similar limb darkening coefficients in the two nearby bandpasses.  Therefore, we adopt an alternative approach.  

For both TrES-2 and TrES-3, we model the LCs using four different scenarios.  First, we fit separate models to the flux ratios at the two wavelengths (columns 1 and 2 of Table \ref{tab_two}), fixing both $c_1$ and $c_2$ at a pair of self-consistent values based on PHOEBE models using a single set of spectroscopic parameters.  The transit and limb darkening parameters in these first two scenarios need not necessarily be self-consistent.  Comparing these first two models allows us to evaluate the sensitivity of our results to the two different bandpasses.  Second, we correct each LC based on the previous results for the fitted slope and mean OOT flux and fit a single model to the the LCs for both bandpasses simultaneously (column 3 of Table \ref{tab_two}).  In this scenario, the transit and planetary parameters ($t_0$, $b$, $D$, $p$) are forced to be the same for both bandpasses (self-consistent if we neglect the possibilities of wavelength-dependent planet radius and contamination from background light).  For this and the subsequent scenario, we fix all four limb darkening parameters (two for each band pass) at self-consistent values based on PHOEBE models.  Finally, we fit a single model to the two (corrected) LCs, but allow for separate values of $D$ and $p$ for each LC (column 4 of Table \ref{tab_two}).  For the final scenario, we continue to force the fit to have the same $t_0$ and $b$ for both bandpasses since these should be the same regardless of the bandpass used for observations, but allow the models in column 4 to have different values of $p$ and $D$ in the two bandpasses so as to account for a potential difference in the planet radius in the two bandpasses or blending of a putative binary star (see \S\ref{SecDiscuss}).  Since we have observations of a single transit in multiple bands and impose additional constraints that all four limb darkening parameters be self-consistent, one would expect our method to result in a larger RMS scatter about the model than the fitting procedures used by previous authors and developed for analyzing transit observations at a single waveband.  

The primary differences in the model parameters among the four scenarios can be traced to different sets of limb darkening parameters.  As \citet{Southworth2008} demonstrated, fixing both LD coefficients at their theoretical values could produce measurement uncertainties that are too small.  To account for the uncertainty in the limb darkening model, we repeat each of the analyses varying the spectroscopic parameters by the published uncertainties and thus the calculated limb darkening coefficients.  We report model parameters for the best-fitting model with the smallest $\chi^2$ value, but present their parameter uncertainties based on the complete set of best-fitting values and uncertainties estimated for all allowed LD parameters (see Table \ref{tab_two}).  For example, we estimate the upper error bar for a given parameter as the result of subtracting its best-fitting value (as given by the model with the smallest $\chi^2$ value) from the maximum sum of a fit value and its associated measurement uncertainty, considering the full set of models computed for the different LD coefficients.  

To further investigate the measurement errors, we applied the ```Prayer Bead'' method \citep[see, e.g.,][]{desert2009}.  Specifically, we construct synthetic light curves by calculating the residuals from the initial best-fitting light curve model, performing a circular shift, and adding the shifted residuals back to the best-fitting light curve model.  Then, we conducted the above analysis on the synthetic data sets.  We use the dispersion of the fit parameters to estimate the effects of any other systematic noise sources not removed by the EPD technique.  We discuss the results from this analysis as well as the formal 1-$\sigma$ errors (as determined from the covariance matrix) for the best-fitting parameters in the following section for the case presented in column 4.

\begin{table*}
 \centering
 \begin{minipage}{140mm}
  \caption{System Parameters of TrES-2 and TrES-3\label{tab_two}}
  \begin{tabular}{@{}ccccc@{}}
  \hline
\multicolumn{1}{c}{Parameter} & \multicolumn{4}{c}{Value} \\
\multicolumn{1}{c}{} & \multicolumn{1}{c}{$\lambda_1$\footnote{We label the bandpass centred on 790.2-nm as $\lambda_1$ for simplicity.}}
 & \multicolumn{1}{c}{$\lambda_2$\footnote{We label the bandpass centred on 794.4-nm as $\lambda_2$ for simplicity.}}
 & \multicolumn{1}{c}{$(\lambda_1+\lambda_2)$\footnote{The best-fitting parameters were determined from the joint analysis of the two light curves for each target, with only the LD coefficients determined separately for the two wavelengths.  See text for additional details.}}
 & \multicolumn{1}{c}{$(\lambda_1+\lambda_2)$\footnote{Same as in column 3, but also with $D$ and $p$ determined separately for each wavelength.}}\\
 \hline
$$ & \multicolumn{4}{c}{TrES-2} \\
\multicolumn{1}{l}{$t_0-$2455000 (HJD$)$}\dotfill & \multicolumn{1}{c}{$7.64601^{+0.00018}_{-0.00021} $} & \multicolumn{1}{c}{$7.64610^{+0.00021}_{-0.00019} $} & \multicolumn{1}{c}{$7.64605^{+0.00013}_{-0.00013} $} & \multicolumn{1}{c}{$7.64605^{+0.00013}_{-0.00013} $} \\
\multicolumn{1}{l}{$b$}\dotfill & \multicolumn{1}{c}{$0.8647^{+0.0051}_{-0.0413}$} & \multicolumn{1}{c}{$0.8638^{+0.0055}_{-0.0335}$} & \multicolumn{1}{c}{$0.8639^{+0.0036}_{-0.0349}$} & \multicolumn{1}{c}{$0.8642^{+0.0036}_{-0.0349}$} \\
\multicolumn{1}{l}{$D$ (days)}\dotfill & \multicolumn{1}{c}{$ 0.07369^{+0.00203}_{-0.00073} $} & \multicolumn{1}{c}{$  0.07478^{+0.00224}_{-0.00078} $} & \multicolumn{1}{c}{$0.07415^{+0.00188}_{-0.00051} $} & \multicolumn{1}{c}{$\lambda_1: 0.07365^{+0.00214}_{-0.00061} $} \\
\multicolumn{1}{l}{}  & \multicolumn{1}{c}{} & \multicolumn{1}{c}{} & \multicolumn{1}{c}{} & \multicolumn{1}{c}{$\lambda_2: 0.07481^{+0.00176}_{-0.00064} $} \\
\multicolumn{1}{l}{$p$}\dotfill & \multicolumn{1}{c}{$0.1254^{+0.0050}_{-0.0009} $} & \multicolumn{1}{c}{$0.1245^{+0.0054}_{-0.0010} $} & \multicolumn{1}{c}{$0.1249^{+0.0048}_{-0.0007} $} & \multicolumn{1}{c}{$\lambda_1: 0.1253^{+0.0052}_{-0.0008} $} \\
\multicolumn{1}{l}{} & \multicolumn{1}{c}{} & \multicolumn{1}{c}{} & \multicolumn{1}{c}{} & \multicolumn{1}{c}{$\lambda_2: 0.1245^{+0.0046}_{-0.0008} $} \\
\multicolumn{1}{l}{$c_1$ (fixed)}\dotfill & \multicolumn{1}{c}{$0.220$} & \multicolumn{1}{c}{$0.232$} & \multicolumn{1}{c}{$\lambda_1: 0.220 $} & \multicolumn{1}{c}{$\lambda_1: 0.220$} \\
\multicolumn{1}{l}{$$} & \multicolumn{1}{c}{} & \multicolumn{1}{c}{} & \multicolumn{1}{c}{$\lambda_2:  0.232$} & \multicolumn{1}{c}{$\lambda_2: 0.232$} \\
\multicolumn{1}{l}{$c_2$ (fixed)}\dotfill & \multicolumn{1}{c}{$-0.084$} & \multicolumn{1}{c}{$-0.086$} & \multicolumn{1}{c}{$\lambda_1: -0.084$} & \multicolumn{1}{c}{$\lambda_1: -0.085$} \\
\multicolumn{1}{l}{$$} & \multicolumn{1}{c}{} & \multicolumn{1}{c}{} & \multicolumn{1}{c}{$\lambda_2: -0.086$} & \multicolumn{1}{c}{$\lambda_2: -0.098$} \\
\multicolumn{5}{c}{ } \\
$$ & \multicolumn{4}{c}{TrES-3} \\
\multicolumn{1}{l}{$t_0-$2455000 (HJD$)$}\dotfill & \multicolumn{1}{c}{$54.52450^{+0.00014}_{-0.00015} $} &  \multicolumn{1}{c}{$54.52445^{+0.00015}_{-0.00015} $} &  \multicolumn{1}{c}{$54.52449^{+0.00012}_{-0.00010} $} & \multicolumn{1}{c}{$54.52447^{+0.00011}_{-0.00011} $} \\
\multicolumn{1}{l}{$b$}\dotfill & \multicolumn{1}{c}{$0.8276^{+0.0575}_{-0.0082} $} & \multicolumn{1}{c}{$0.8737^{+0.0178}_{-0.0453} $} & \multicolumn{1}{c}{$0.8716^{+0.0123}_{-0.0455} $} & \multicolumn{1}{c}{$0.8356^{+0.0482}_{-0.0073} $} \\
\multicolumn{1}{l}{$D$ (days)}\dotfill & \multicolumn{1}{c}{$0.05779^{+0.00080}_{-0.00194} $} & \multicolumn{1}{c}{$0.05554^{+0.00202}_{-0.00086} $} & \multicolumn{1}{c}{$0.05608^{+0.00160}_{-0.00061} $} & \multicolumn{1}{c}{$\lambda_1: 0.05817^{+0.00072}_{-0.00203} $} \\
\multicolumn{1}{l}{} & \multicolumn{1}{c}{} & \multicolumn{1}{c}{} & \multicolumn{1}{c}{} & \multicolumn{1}{c}{$\lambda_2: 0.05652^{+0.00069}_{-0.00174} $} \\
\multicolumn{1}{l}{$p$}\dotfill & \multicolumn{1}{c}{$0.1672^{+0.0071}_{-0.0052} $} & \multicolumn{1}{c}{$0.1641^{+0.0067}_{-0.0067} $} & \multicolumn{1}{c}{$0.1662^{+0.0046}_{-0.0048} $} & \multicolumn{1}{c}{$\lambda_1: 0.1695^{+0.0046}_{-0.0049} $} \\
\multicolumn{1}{l}{} & \multicolumn{1}{c}{} & \multicolumn{1}{c}{} & \multicolumn{1}{c}{} & \multicolumn{1}{c}{$\lambda_2: 0.1631^{+0.0048}_{-0.0044} $} \\
\multicolumn{1}{l}{$c_1$ (fixed)}\dotfill & \multicolumn{1}{c}{$0.653$} & \multicolumn{1}{c}{$0.222$} & \multicolumn{1}{c}{$\lambda_1: 0.220$} & \multicolumn{1}{c}{$\lambda_1: 0.653$} \\
\multicolumn{1}{l}{} & \multicolumn{1}{c}{} & \multicolumn{1}{c}{} & \multicolumn{1}{c}{$\lambda_2: 0.222$} & \multicolumn{1}{c}{$\lambda_2: 0.625$} \\
\multicolumn{1}{l}{$c_2$ (fixed)}\dotfill & \multicolumn{1}{c}{$0.263$} & \multicolumn{1}{c}{$$-0.086} & \multicolumn{1}{c}{$\lambda_1: -0.081$} & \multicolumn{1}{c}{$\lambda_1: 0.263$} \\
\multicolumn{1}{l}{} & \multicolumn{1}{c}{} & \multicolumn{1}{c}{} & \multicolumn{1}{c}{$\lambda_2: -0.086$} & \multicolumn{1}{c}{$\lambda_2: 0.233$} \\
\hline
\end{tabular}
\end{minipage}
\end{table*}

\section{Results}
\label{SecResults}

We show the flux ratios and the best-fitting models (after correcting for slopes/trends based on results given in column 4 of Table \ref{tab_two}) in Figures \ref{FigTres2} and \ref{FigTres3}.  Note that the time series given in Table \ref{tab_one} was corrected based on results from column 4 of Table \ref{tab_two} as well.  The best-fitting LC parameters for TrES-2 and TrES-3 are given in Table \ref{tab_two} and described in \S\ref{SecPlanets}.  We analyze the residuals and discuss the photometric precision in
\S\ref{SecBins}.

\subsection{Planetary Parameters}
\label{SecPlanets}

For the joint analysis presented in column 4 of Table \ref{tab_two} (which we adopt as our baseline model), the best-fitting parameter values and their formal 1-$\sigma$ uncertainties for TrES-2 are: $t_0=2455007.64605\pm0.00013$ (HJD), $b=0.8642\pm0.0036$, $D$ (790.2nm) $=0.07365\pm0.00061$ days, $D$ (794.4nm) $=0.07481\pm0.00063$ days, $p$ (790.2nm) $=0.1253\pm0.0008$, and $p$ (794.4nm) $=0.1245\pm0.0008$.  Similarly, for the joint analysis of TrES-3, the best-fitting parameter values and their formal 1-$\sigma$ uncertainties are: $t_0=2455054.52447\pm0.00010$ (HJD), $b=0.8356\pm0.0073$, $D$ (790.2nm) $=0.05817\pm0.00065$ days, $D$ (794.4nm) $=0.05652\pm0.00069$ days, $p$ (790.2nm) $=0.1695\pm0.0025$, and $p$ (794.4nm) $=0.1631\pm0.0022$.  We note that for both TrES-2 and TrES-3, these uncertainties are typically smaller than those in Table \ref{tab_two}, since those also account for uncertainty in the LD model.  Additionally, the 1-$\sigma$ errors for the median best-fitting parameters from the Prayer Bead analysis for both TrES-2 and TrES-3 are slightly larger than, but still comparable to, the formal 1-$\sigma$ uncertainties, with the largest deviations occurring for the uncertainty in the radius ratio.  We also computed the RMS of the best-fitting parameter values from the Prayer Bead analysis as an additional check on our measurement errors.  We find that accounting for the uncertainty in the LD model sufficiently accounts for the distribution of the best-fitting parameters as derived from the Prayer Bead analysis.  

For both TrES-2 and TrES-3, the best-fitting values for each parameter are consistent between the different models presented in Table \ref{tab_two}, but only after accounting for uncertainties in the limb darkening model.  The most notable difference in our results with different modeling procedures is for TrES-3 when comparing the impact parameter in columns 2 and 3 (analysis for the 794.4-nm light curve and the joint analysis with common planet parameters) with the other two models for TrES-3.  This is at least partly due to the near degeneracy between impact parameter and limb darkening model.  Indeed, the LD coefficients for the models in column 2 and 3 are significantly different from the best values for the other two cases.  While we were able to place tight constraints on the impact parameter for TrES-2b, the precision for TrES-3b is significantly reduced, since we are modeling an incomplete light curve and have no data prior to ingress.

The best-fitting slopes and their formal uncertainties as measured in the individual LCs before performing the joint analyses for the TrES-2 and TrES-3 LCs are $0.00012\pm0.00172$ and $-0.00020\pm0.00180$ days$^{-1}$ for the 790.2 and 794.4 nm LCs for TrES-2, and $0.00256\pm0.00501$ and $0.00092\pm0.00540$ days$^{-1}$ for TrES-3.  While the estimated slopes for TrES-2 are fairly consistent, the difference in the estimated slopes is much larger for TrES-3 than for TrES-2.  Note that the measured slopes account for the effects of both differential atmospheric extinction (due to some reference stars being observed at a different wavelength than the target and primary reference) and real astrophysical variability in the colour of the target and/or reference stars.  Our observations are not generally able to disentangle the two potential causes of a slope.  If the slope were primarily due to differential extinction, then observations of multiple transits should produce consistent results.  On the other hand, if stellar variability is significant, then each transit would need a separate slope parameter.  In the case of TrES-3, the large difference in slopes between the two bandpasses suggests that differential extinction is not responsible.  However, we caution that the measurement of the slope around transit is likely affected by the lack of pre-transit data.  Thus, we emphasize the importance of acquiring complete transit LCs as well as extended, uninterrupted OOT baseline data.  

For TrES-2, the uncertainties in the OOT flux (as determined from the individual LC analyses) were $\sim0.000091$ ($0.000097$) for the 790.2 (794.4) nm observations, based on 26 (25) exposures totaling $\sim$35 (33) minutes of integration time.  The uncertainties for the OOT flux for TrES-3 are slightly larger, perhaps due to the lack of pre-transit observations and covariance with the slope.  However, the TrES-3 observations were more stable overall and allowed for a single baseline of OOT data that was longer than either the pre- or post-transit baselines for the TrES-2 observations.  Both observations demonstrate the capability for very high-precision measurements that could enable the detection of super-Earth-sized planets around solar-like stars and/or the characterization of the atmospheres of giant planets orbiting bright stars (see \S\ref{SecBins} \& \ref{SecDiscuss}).

\subsection{Light Curve Residuals}
\label{SecBins}

We computed the residual flux by subtracting the best-fitting models given by the joint analysis presented in column 4 of Table \ref{tab_two} from the data (see Figures \ref{FigTres2}b,c and \ref{FigTres3}b,c).  Note that some of the residuals show evidence of additional systematic noise sources that were not removed with the EPD technique.  While we do not know the origin of these systematics, we note that the uncertainty estimates from the Prayer Bead analysis are consistent with our best-fitting parameters (as discussed above, in \S\ref{SecPlanets}).  

We estimate the photometric precision to be 343$\pm$45 and 412$\pm$43 parts per million (ppm) for observations of TrES-2 at 790.2 and 794.4 nm, based on the root mean square deviation of the residuals.  Despite high airmass (up to $\sim$2.22), the TrES-3 observations produced a precision of 470$\pm$64 and 424$\pm$57 ppm at 790.2 and 794.4 nm.  Because the estimated precisions for both targets are smaller than (or comparable to, based on the upper limits computed from the expected standard deviations) the estimated measurement uncertainties (see \S\ref{SecReduction}), we take them to be consistent with the theoretical limit.  

\begin{figure}
\includegraphics[width=84mm]{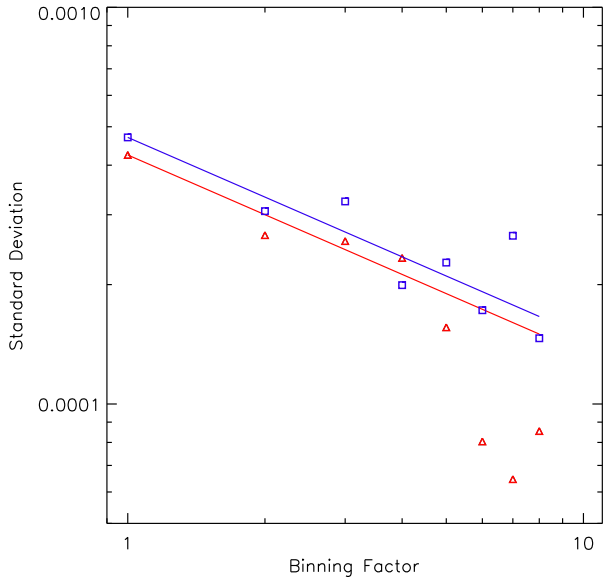} 
\caption{Standard deviation of time-binned residuals for TrES-3 as a function of the number of data points per bin ($n$).  The blue square (red triangle) symbols are the standard deviations of the 790.2 (794.4) nm binned light curve residuals.  The solid curves show the trend expected for white Gaussian noise ($n^{1/2}$).  Each exposure was separated by 5.11 minutes, thus the present observations only allow us to test timescales of up to $\sim$40 minutes due to the limited duration of the observations.}
\label{FigBins}
\end{figure}

Due to the long, uninterrupted OOT baseline obtained for TrES-3, we take the TrES-3 observations as the best-case scenario to achieve high precisions consistent with the theoretical limit and present the standard deviation of the time-binned residuals for the TrES-3 LCs in Figure \ref{FigBins}.  The residuals for the 790.2-nm LC are consistent with the trend expected for white Gaussian noise, while the residuals for the 794.4-nm LC deviate at binning factors larger than 5.  When binning the uncorrelated 790.2-nm residuals over longer observing times ($\sim$40 min), we estimate a precision of 146 ppm, which is sufficient to detect the transit of a super-Earth-size planet ($\ge~1.3 R_{\oplus}$) orbiting a solar-like star or an Earth-size planet orbiting a star smaller than 0.7 $R_\odot$ (assuming $\sim$120 min of observations between 2nd and 3rd contact and 3-$\sigma$ confidence level).

\subsection{Transit Colour}
\label{SecColour}

In Figures \ref{FigTres2}d and \ref{FigTres3}d, we present the colour of the residual fluxes.  The colour was computed by averaging each pair of points in the 790.2-nm LC residuals and dividing by the flux residuals of the 794.4-nm LC.  We find no significant colour deviation in either of the TrES-2 or TrES-3 systems.  We estimate RMS precisions of 519 ppm and 502 ppm for TrES-2 and TrES-3, which are both slightly larger, but still consistent, with the precisions estimated for the individual LCs.  We found no significant difference in the colours in and out of transit for TrES-2.  For TrES-3, there was a slight slope in the colour during transit, but we do not consider this to be significant given the large uncertainty in the slopes fit to the individual LCs due to the lack of pre-transit observations. 

\section{Discussion}
\label{SecDiscuss}

The first exoplanet observations from the GTC provided excellent photometric precision, despite problems with the telescope, high airmass and relatively poor atmospheric conditions.  

Our results for TrES-3 are consistent with previous high-precision observations \citep{gibson2009,sozzetti2009}.  Our measurements of planet-star radius ratio, impact parameter and transit duration bracket those of previous studies.  Our measured transit time has an uncertainty of $\sim$9.5-s and occurs 44-s after that predicted by \citet{sozzetti2009} and 14-s after the updated ephemeris of \citet{gibson2009}.  \citet{gibson2009} conducted a transit timing analysis of TrES-3b and they find no evidence of transit timing variations.  Given the differences between various ephemerides and the possibility for stellar variability to contribute additional timing noise, we do not consider such a difference significant.  

Our results for for TrES-2b are also in good agreement with \citet{holman2007} and \citet{sozzetti2007}.  However, we note that all reported values for the radius ratio could be affected by a binary companion to TrES-2, which has a separation of 1.089 $\pm$ 0.008 arcsec from the primary star and $\Delta~i'\simeq~3.7$mag as reported by \citet{daemgen2009}.  While the companion star is clearly included within our aperture, our results do not account for the flux of the companion in our analysis.  In this case, our model parameter $p$ serves as a depth parameter which can be related to the actual planet-star radius ratio for a given amount of contamination \citep{daemgen2009}.  If we assume that the primary-secondary star flux ratio in our bandpasses is similar to that in $i'$, then the planet-star radius ratio would increase by $\sim$1.6\% relative to our estimates in Table \ref{tab_two}.  In principle, the different colour of a blended star could result in different transit depths when observed in multiple different bandpasses.  While our results are suggestive of such a transit depth difference, the difference is not statistically significant.  Given the high-precision of our observations and a possible feature in the in-transit data (around mid-transit), we suspect that the apparent differences in $p$ may be due to variations in the stellar surface brightness.\footnote[5]{Indeed, previous studies have found other indications of possible stellar activity, including variations in the OOT flux \citep{odonovan2006,raetz2009}.  Our observations show no signs of a putative ``second dip'' \citep{raetz2009}, but only extend for $\sim$1 hour after the end of transit.}  In light of these potential complications, near-simultaneous photometry in multiple bandpasses could be particularly useful for recognizing when a companion or background object is affecting the photometry.

Several recent studies have considered the possibility of variations in the transit times and/or durations of TrES-2b \citep{mislis2009,mislisetal2009,rabus2009,raetz2009,scuderi2009}.  Our best-fitting transit time has a measurement uncertainty of $\sim$11-s and is offset from the predicted ephemeris of \citet{holman2007} by $\sim$4 minutes, \citet{rabus2009} by $\sim$142-s, \citet{raetz2009} by $\sim$16-s and \citet{scuderi2009} by $\sim$64-s.  Given the differences in various ephemerides, observing bandpasses and limb darkening models, we do not consider the transit time offset to be significant evidence for transit timing variations.  This is consistent with the recent transit timing analysis of \citet{raetz2009}.  While \citet{rabus2009} raised the possibility of sinusoidal variations in the transit ephemeris due to an exomoon, they do not find results of the magnitude proposed by \citet{mislis2009}.  

Due to TrES-2b's large impact parameter, the transit duration is very sensitive to changes induced by additional planets or other bodies in the system \citep{miralda2002}.  \citet{mislis2009} and \citet{mislisetal2009} claim that the transit duration decreased by three minutes between 2006 and 2008, and they argue that a third body is a natural explanation for this change.  If we include uncertainty due to unknown limb darkening models, we cannot definitively rule out the shorter durations suggested by \citet{mislis2009} and \citet{mislisetal2009}.  When interpreting putative differences in transit duration/impact parameter, one should be mindful of differences in the various filters and the stellar LD model used \citep{scuderi2009}.  Indeed, \citet{scuderi2009} report no significant change in the orbital inclination/transit duration when they compare their observations with those of \citet{odonovan2006} (taken in the same filter).  Since our observations use unique bandpasses, it is not (yet) possible to compare them to previous observations with the same bandpass.  Since TrES-2 is in the {\em Kepler} field, {\em Kepler} observations should soon shed light on this matter, as the observations will cover longer timescales and therefore multiple transits all within the same passband.

In summary, we find that large ground-based observatories are capable of achieving high-precision differential photometry.  Systematic effects due to variable atmospheric extinction can be minimized by the use of a small far-red or NIR bandpass chosen to avoid sky absorption lines.  Thus, large, ground-based observatories can help characterize super-Earth-size planets discovered by ongoing transit searches, including planets in the habitable zone of main-sequence stars for which transits last several hours.  We anticipate that this innovative technique for high-precision photometry will enhance the ability of current and future large optical/NIR observatories to study the properties of both Earth-like planets (e.g., size, density and orbit) and giant planets.  For example, the high photometric precision could allow for the detection of the occultation of hot transiting giant planets like TrES-3b at optical wavelengths.  While the occultation depth probes the temperature and dynamics of the planet's atmosphere, the time and duration of occultation provide a powerful probe of the orbit's eccentricity.

The technique is even more powerful when observations in multiple bandpasses can be obtained nearly simultaneously and thus during the same transit.  Such observations could help verify that the transit of a candidate planet is not due to stellar variability plus rotation by observing in multiple wavelengths that reduce the effects of stellar variability and/or enhance the contrast of spots or plage.  Similarly, nearly simultaneous observations in multiple narrow bandpasses could be used to characterize the atmospheres of giant planets.  For example, HST and HET have used transmission spectroscopy of HD 209458b and HD 189733b to measure Na I absorption in the planetary atmospheres \citep{charb2002,redfield2008}.  Future GTC observations could be used to perform similar measurements of atmospheric features such as Na I and K I absorption in planets transiting bright stars, especially once planned improvements in the OSIRIS CCD controller software reduce the dead time between exposures.  

Considering the increasing numbers of known transiting exoplanets, we hope that this novel technique will be an additional useful tool to help characterize a variety of planetary properties.

\section*{Acknowledgments }

We gratefully acknowledge the observing staff at the GTC and give a special thanks to Ren\'e Rutten, Antonio Cabrera Lavers, Jos\'e Miguel Gonz\'alez, Jordi Cepa Nogu\'e and Daniel Reverte for helping us plan and conduct these observations successfully.  We thank Andrej Prsa for computing limb darkening models for our unique bandpasses.  We appreciate early feedback from Dave Charbonneau, Scott Gaudi, Matthew Holman and Heather Knutson.  We also thank the anonymous referee for helping us improve this manuscript.  K.D.C.\ is supported by an NSF Graduate Research Fellowship.  C.H.B. received support from an NSF Astronomy and Astrophysics Postdoctoral Fellowship.  This work is based on observations made with the Gran Telescopio Canarias (GTC), installed in the Spanish Observatorio del Roque de los Muchachos of the Instituto de Astrof\'isica de Canarias, on the island of La Palma.  The GTC is a joint initiative of Spain (led by the Instituto de Astrof\'isica de Canarias), the University of Florida and Mexico, including the Instituto de Astronom\'ia de la Universidad Nacional Aut\'onoma de M\'exico (IA-UNAM) and Instituto Nacional de Astrof\'isica, \'Optica y Electr\'onica (INAOE).

\label{lastpage}

\end{document}